# Broadband and wide-temperature-range thermal emitter with super-hydrophobicity based on oxidized high-entropy film

Ping Song[1], Cong Wang[1,*], Ying Sun[1], Angélique Bousquet[2], Eric Tomasella[2]

[1] *Center for Condensed Matter and Material Physics, Beihang University, Beijing 100083, PR China*

[2] *Université Clermont Auvergne, CNRS, SIGMA Clermont, Institut de Chimie de Clermont-Ferrand (ICCF), Clermont-Ferrand F-63000, France*

**Abstract**

This paper demonstrates broadband and wide-temperature-range thermal emitter with super-hydrophobicity based on an oxidized high-entropy film. The annealed (NiCuCrFeSi)O high-entropy film can serve as a mid-infrared emitter due to its three characteristics: i) High blackbody-like emission in a broadband mid-infrared region from 2.5 $\mu$m to 20 $\mu$m; ii) High emittance (more than 90%) in a wide temperature range from 300 K to 1000 K; iii) Super-hydrophobicity with high contact angle (CA) of 151° and low title angle (TA) of 5°. These characteristics originate from the recrystallization of as-deposited film and formation of cavernous structure with a plethora of nanoscale polyhedrons and holes after an annealing process, supported by the results of XRD and SEM/EDX. This work may provide new ideas for applications of thermal emitters operating at substantially wide temperature range.

## I. INTRODUCTION

Blackbody radiation follows Planck's law that the emission profile is related to the wavelength of electromagnetic radiation and shifts with temperature [1,2]. Thermal emitters based on the blackbody radiation are widely used but not limited in applications of energy generation (thermophotovoltaic devices [3–6] and thermoelectric devices [7–9]), radiative cooling [10,11], sensing [12,13], spectroscopy [14,15], etc. A perfect thermal emitter requires a broadband emission spectrum and its thermal emission ability is strongly dependent on the structure and materials [16]. There are a lot of state-of-the-art works on designing of perfect thermal emitters. Shi *et al.* [17] realized a broadband thermal emission with average emittance is more than 70 % by using a hexagonal boron nitride-

---

encapsulated graphene filament. Lenert et al. [18] reported a high-efficiency solar thermophotovoltaic device coupling with a one-dimensional Si/SiO$_2$ emitter with thermal efficiency is ~ 75%. Shi et al. [19] theoretically performed 1.87- and 2.94-fold better in thermal radiation based on 2D materials. Raman et al. [11] experimentally demonstrated a multilayer thermal emitter to realize a passive radiative cooling to ~ 5 °C below the ambient air temperature under direct sunlight. Rephaeli et al. [10] achieved a high-performance daytime radiative cooling (net cooling power > 100 W/m$^2$ at ambient temperature) by designing a broadband-photonic-structure thermal emitter. Hossain et al. [20] fabricated a metamaterial-structure thermal emitter in the wavelength range from 2 $\mu$m to 20 $\mu$m to cool down 12.2 °C below the ambient temperature. However, all these devices are performed as metasurfaces, photonic crystals, or nanoscale multilayers, which must go through a complex manufacturing process in a limited size.

In this paper, we report a perfect thermal emitter based on a single-layer oxidized high-entropy film after the annealing process. Three characteristics of this emitter are observed: i) High blackbody-like emission in a broadband mid-infrared region; ii) High emittance (more than 90%) in a wide temperature range from 300 K to 1000 K; iii) Super-hydrophobicity.

## II. EXPERIMENTAL DETAILS

The (NiCuCrFeSi)O (referred to below as (NCCFS)O) thin films were deposited on silicon and stainless steel (SS) substrates in a magnetron sputtering system at room temperature. The base pressure of the chamber was kept at ~ $5 \times 10^{-7}$ Pa before deposition, and the working pressure was kept at 0.5 Pa. The 3-inch alloy target was synthesized by powder alloy method with an equiatomic composition of pure metals of Ni, Cr, Cu, and Fe (> 99.99 wt. %) and another 3-inch target is Si (> 99.99 wt. %). Both targets were placed at the tilt angle of 60° in the vacuum chamber toward the center of the rotating substrate. The argon gas flow rate was kept at 6 sccm, and the oxygen gas flow rate was kept at 0.5 sccm. The target power density on the alloy target was fixed at 2.2 W/cm$^2$ and the power density on the Si target was fixed at 4.4 W/cm$^2$. The as-deposited films then were annealed at 1273 K for 2 h and cooled slowly to room temperature in vacuum.

X-ray diffraction (XRD) patterns with Cu-$K\alpha$ radiation were obtained in a 2$\theta$ range of 10° – 70° at room temperature. The thickness of the as-deposited film was confirmed by a cross-sectional scanning electron microscope (SEM) image. Composition and homogeneity of films were evaluated by energy-dispersive X-ray (EDX) analysis combined on SEM and Rutherford backscattering spectroscopy (RBS) (Orléans, France). The reflectance ($R$) spectra were measured by a Fourier transform infrared (FTIR) spectrophotometer. The wettability of the film surface was investigated by the water contact angle (CA) and title angle (TA) measurement.

## III. RESULTS AND DISCUSSION

### A. Determination of film structure and composition before and after an annealing process

The crystal structure of the (NCCFS)O films and SS substrates before and after annealing at $T = 1273$ K in vacuum is confirmed by XRD patterns. As shown in Figure 1, for the XRD pattern of the as-deposited (NCCFS)O film (blue line), there are two main peaks at the positions of $2\theta = 43.6°$ and $2\theta = 50.7°$ which are confirmed as the (111) peak and (002) peak, respectively, of the SS substrate (black line). There are no more characteristic peaks observed except for a small shoulder around the (111) peak, indicating an amorphous structure of the as-deposited (NCCFS)O film. After an annealing process from $T = 1273$ K in vacuum, the XRD pattern (yellow line) of the (NCCFS)O film is different from that of the as-deposited film. There are six new peaks observed at the positions of $2\theta = 18.2°$, 21.9°, 29.9°, 35.2°, 56.6°, and 62.1°, respectively. This could be caused by the recrystallization of the amorphous film or oxidation of the film and substrate during the thermal treatment. However, if we check the XRD pattern (red line) of the SS substrate after the same annealing process, there are no more additional peaks observed besides two main SS characteristic peaks. Therefore, we believe that the appearance of these new characteristic peaks is due to recrystallization. The new peaks are indexed as the (111), (002), (022), (113), (115), and (044) peaks, respectively, indicating the face-centered cubic (fcc) structure of the annealed (NCCFS)O film.

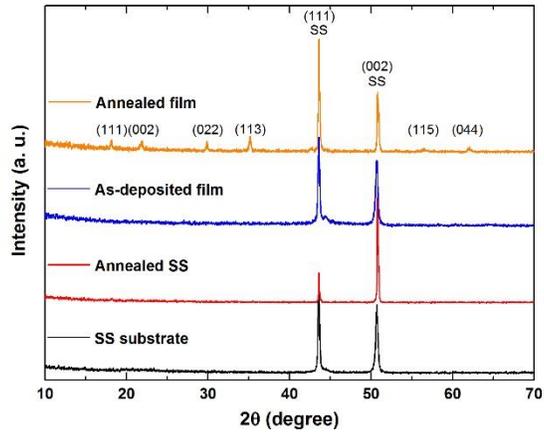

Fig. 1. XRD patterns of the (NCCFS)O films and substrates before and after annealing at $T$ = 1273 K in vacuum.

The microstructure and elemental distribution of the as-deposited (NCCFS)O film are shown in Fig. 2. Fig. 2a shows the top view SEM image of the film surface with the scale bar of 2 μm, indicating a smooth surface of the as-deposited (NCCFS)O film. SEM/EDX mapping, as shown in Figs. 2(b−g), reveals the homogeneous distribution of all elements. The elemental ratios of the as-deposited film confirmed by EDX are listed in Table 1. It can be seen that the five elements have a near-equal elemental ratio with the concentrations of each of them being between 5% and 35% except for oxygen atom. We hence consider the as-deposited (NCCFS)O film as a high-entropy film [21].

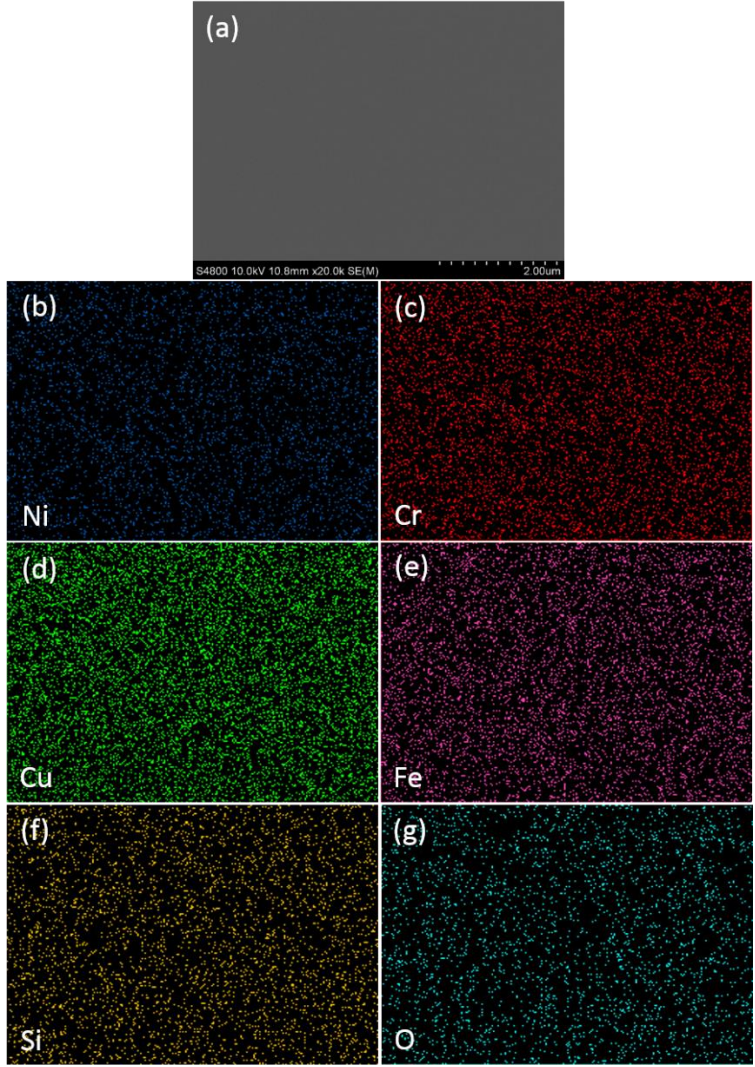

Fig. 2. Microstructure and elemental distribution of the as-deposited (NCCFS)O film with the scale bar of 2 μm. (a) SEM image of the film surface. Distribution map of (b) Ni; (c) Cr; (d) Cu; (e) Fe; (f) Si; (g) O, measured by SEM/EDX.

**Table 1**
Elemental ratios of the (NCCFS)O film confirmed by EDX and RBS before and after annealing at $T = 1273$ K in vacuum.

|  |  | Ni | Cr | Cu | Fe | Si | O |
|---|---|---|---|---|---|---|---|
| As-deposited film (%) | EDX | 12.69 | 15.65 | 29.01 | 21.27 | 17.17 | 4.21 |
|  | RBS | 14.02 | 17.82 | 22.31 | 15.76 | 22.95 | 7.14 |
| Annealed film (%) | EDX | 6.30 | 6.56 | 14.46 | 35.16 | 5.41 | 32.11 |

Much more detailed analysis of the elemental composition and thickness of the as-deposited film is carried out by RBS and cross-sectional SEM. The RBS has been traditionally used to determine the structure and composition of materials by measuring the backscattering of a beam of high energy ions (2000 keV $^4$He ions in our case) impinging on a sample and its main advantage is fully quantitative [22–24]. Fig. 3a shows the RBS result of the as-deposited (NCCFS)O film on the silicon substrate. By fitting the experimental data (black cross), the elemental composition of the as-deposited film can be quantitatively estimated. The fitting result (solid red line) shows a good agreement with the experimental data. The channel positions of O, Si, Cr, Fe, Ni, and Cu correspond to the remarkable shoulders at the channels of 373, 587, 743, 761, 779, and 812, respectively. The elemental ratios of the as-deposited film confirmed by RBS are shown in Table 1. Compared to the EDX result, the increase of the Si and O contents may be due to the fact that RBS has a different sensitivity to light elements [25]. In addition, the peak around the channel of 756 is symmetric along the central line, indicating the fairly uniform mixture of the elements in the as-deposited film. This is consistent with the EDX results in Fig. 2. Fig. 3b shows the cross-sectional SEM image of the as-deposited (NCCFS)O film. The image once again shows a smooth surface and a clear interface between film and silicon substrate with a film thickness of 320 nm.

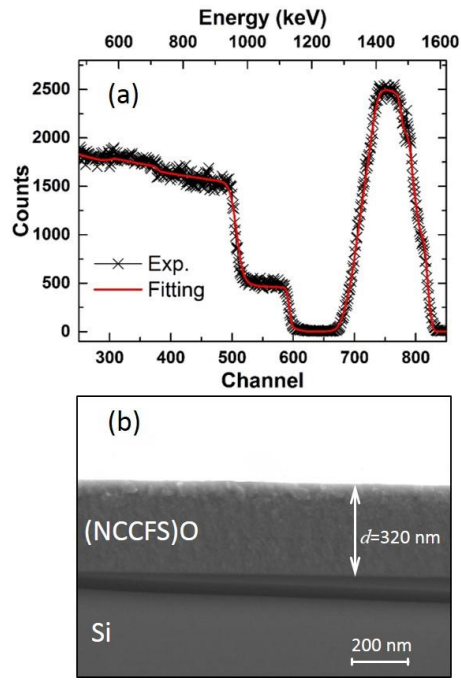

Fig. 3. Determination of the as-deposited film composition by Rutherford backscattering spectroscopy (RBS). (a) RBS spectrum of 2000 keV $^4$He ions from the (NCCFS)O film deposited on Si substrate, scattering angle $\alpha$

= 165°. Black cross: experimental data. Red solid line: fitting result. (b) Cross-sectional SEM image of the (NCCFS)O film deposited on a Si substrate.

The XRD result in Fig. 1 has confirmed an fcc structure of the annealed (NCCFS)O film due to the recrystallization when being annealed at $T = 1273$ K in vacuum. We would like to know more about the elemental distribution and microstructure of the annealed film. Fig. 4a shows the EDX spectrum of the annealed film and the corresponding elemental ratios of the annealed film which are shown in Table 1. Figs. 4(b–d) show the microstructure of the annealed film. The smooth surface before the annealing process transforms into a cavernous structure with a plethora of nanoscale polyhedrons and holes after the annealing process. There are two characteristics of the cavernous structure: i) The polyhedrons and holes have a subwavelength size varies from ~ 0.2 μm to ~ 10 μm; ii) The polyhedrons look more angular with sharp and clear edges. As shown in Fig. 4d, the edges of the polyhedrons are highlighted by green lines. The inset between Figs. 4a and b is the top view of the annealed (NCCFS)O film which shows a black surface and hydrophobic property of the annealed film. Compared to the elemental ratios of the as-deposited film, the contents of O and Fe elements increase significantly. The increase of oxygen content may originate from the air that fills the holes inside the annealed film, and the increase of iron content may come from the uncovered SS substrate under the cavernous-structure annealed film.

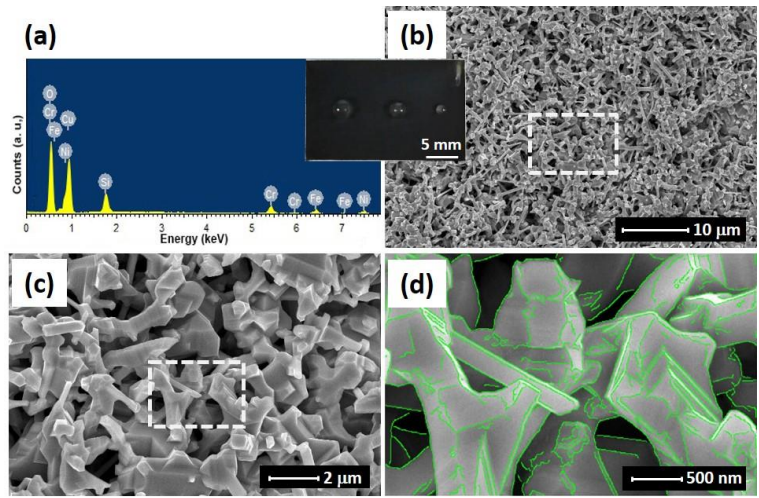

Fig. 4. Elemental determination and microstructure of the (NCCFS)O film annealed at $T = 1273$ K in vacuum. (a) EDX spectrum of the annealed (NCCFS)O film. Inset between (a) and (b) figures is the top view of the annealed (NCCFS)O film with three millimetric water droplets on the surface. (b) SEM image of the annealed (NCCFS)O film. (c) Partially enlarged view of the dash-line marked area in (b). (d) Partially enlarged view of the dash-line marked area in (c). Green lines highlight the edges of the polyhedrons.

### B. Broadband thermal emission of annealed (NCCFS)O film

To investigate the emission properties of the (NCCFS)O films before and after the annealing at $T = 1273$ K in vacuum, Fig. 5a shows the reflectance ($R$) spectra of the (NCCFS)O films and substrates. Fig. 5b shows the corresponding absorptance ($A$) spectra calculated by $A = 1 - R$ [26]. The absorptance of the annealed (NCCFS)O film shows a distinctive property compared with that of the as-deposited film and SS substrates before and after the annealing process. As shown in Fig. 5b, the absorptance of the annealed film is almost more than 90 % over a very broad range of wavelengths, effectively spanning from $\lambda_{min} = 2.5$ $\mu$m to $\lambda_{max} = 20$ $\mu$m. This may be due to the "light-trapped" ability of the cavernous-structure polyhedrons (as shown in Fig. 4d) on the surface of the annealed film [27].

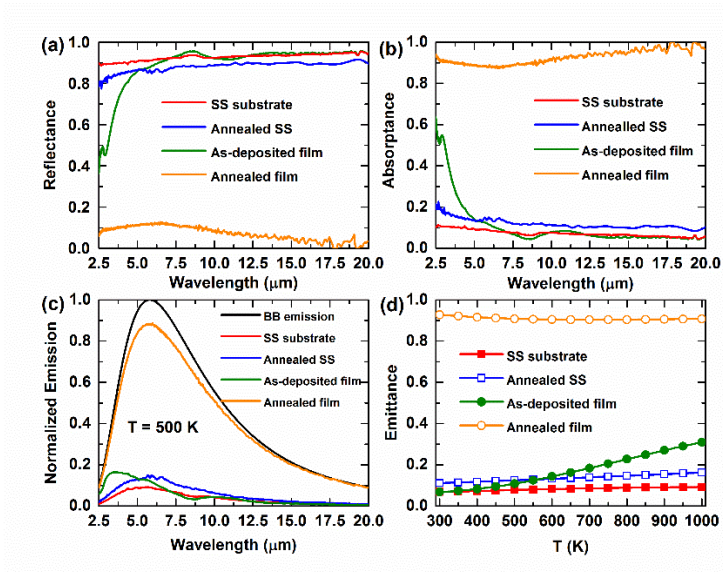

Fig. 5. Comparison of emission properties for the (NCCFS)O films and substrates before and after annealing at $T = 1273$ K in vacuum. (a) Reflectance spectra of the (NCCFS)O films and substrates. (b) Absorptance spectra of the (NCCFS)O films and substrates. (c) Normalized emission of the (NCCFS)O films and substrates compared with the blackbody (BB) emission at $T = 500$ K. (d) Integrated emittance of the (NCCFS)O films and substrates as a function of temperature.

The emission ($E$) is easily derived from Kirchoff's law of thermal radiation [28]:

$$E = AI_b(\lambda, T) = (1-R)I_b(\lambda, T) \quad (1)$$

where $I_b(\lambda, T)$ is the Planck's law distribution of blackbody (BB) radiation at temperature $T$ [29]:

$$I_b(\lambda, T) = \frac{2hc^2}{\lambda^5(\exp(hc/\lambda k_B T)-1)} \quad (2)$$

where $\lambda$ is the wavelength, $h$ the Planck's constant, and $k_B$ the Boltzmann constant. The integrated emittance ($\varepsilon$) at different temperatures can be calculated by integrating wavelength [30]:

$$\varepsilon(T) = \frac{\int_{\lambda_1}^{\lambda_2} (1-R) I_b(\lambda, T) d\lambda}{\int_{\lambda_1}^{\lambda_2} I_b(\lambda, T) d\lambda} \qquad (3)$$

Fig. 5c compares the wavelength dependence of the emission for the (NCCFS)O films and substrates at $T$ = 500 K. It has a high emission for the annealed film compared to that of the as-deposited film and substrates. This blackbody-like emission ensures that the annealed (NCCFS)O film can be operated as a broadband thermal emitter. Fig. 5d shows the integrated emittance of the (NCCFS)O films and substrates as a function of temperature range from 300 K to 1000 K. The emittance of the annealed (NCCFS)O film is distinctively high over the whole temperature range of 300 K – 1000 K, in stark contrast to the results of the as-deposited film and substrates. According to Eqs. 1 and 3, the emission is proportional to the absorption. In our case, the cavernous structure of the annealed film has a subwavelength size and many clear-edge polyhedrons. This special structure gives the annealed film a strong ability to trap the mid-infrared light, i.e. high absorption of the mid-infrared radiation. As a result, the film shows a high emission in the mid-infrared wavelength region.

We conclude that the annealed (NCCFS)O film shows two extraordinary properties in thermal emission: i) High blackbody-like emission in a broadband mid-infrared region; ii) High emittance in a wide temperature range. These two characteristics make the annealed film capable as an emitter in some potential application areas, such as thermophotovoltaic devices [18], radiative cooling [31], and mid-infrared source of thermoelectric devices [9].

### C. Super-hydrophobicity of annealed (NCCFS)O film

The wettability of the annealed (NCCFS)O film was investigated by a water contact angle (CA) and title angle (TA) measurement. Fig. 6 shows the front view of the annealed (NCCFS)O film with three millimetric water droplets (with volumes of 1, 3 and 5 microliters, respectively) at the surface. Further CA image measurement (right inset of Fig. 6) shows a CA $\varphi$ = 151° of the annealed film surface, indicating the *super-hydrophobicity* of the annealed film [32]. As a comparison, the as-deposited film shows hydrophilicity with CA $\varphi$ = 96° (left inset of Fig. 6). The detailed comparison of the dynamic wettability is shown in Videos 1 and 2 as the supplementary materials. The super-hydrophobicity of the annealed film may originate from the Wenzel or Cassie effects [33] of the cavernous structure with a plethora of nanoscale polyhedrons and holes (as shown in Fig. 4d). In general, the Wenzel model assumes a wet contact between water droplets and substrates and the space between the nanoscale polyhedrons at the surface is filled by the liquid, resulting in a high CA hysteresis; the Cassie model assumes a non wet contact between water droplets and substrates due to air trapped between the nanoscale polyhedrons at the surface. So the water droplets can roll off easily.

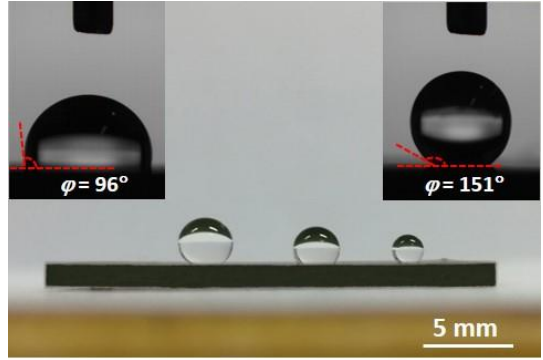

Fig. 6. Front view of the three millimetric water droplets on the surface of the (NCCFS)O film annealed at $T = 1273$ K in vacuum. Right inset shows water CA image of the annealed (NCCFS)O film surface with CA $\varphi = 151°$. Left inset shows water CA image of the as-deposited (NCCFS)O film surface with CA $\varphi = 96°$ for contrast.

To further distinguish the mechanism of hydrophobicity (Wenzel model or Cassie model) in our case, the TA $\theta$ of the annealed (NCCFS)O film is measured by a gradienter fixed on the angle-adjustable inclination platform. Fig. 7a shows the schematic of the experimentation. The TA image of the dynamic rolling process is performed by a camera in front of the platform. Figs. 7(b–d) show the dynamic rolling process of water droplets onto the super-hydrophobic film surface with $\theta = 1°$, $3°$, and $5°$, respectively. It can be seen from Fig. 7b that the water droplet cannot fall off from the film surface when $\theta = 1°$, and only a slight displacement when $\theta = 3°$ (as shown in Fig. 7c). When $\theta = 5°$, the water droplet can easily fall off from the super-hydrophobic film surface (as shown in Fig. 7d). We hence confirm that the TA value of the annealed (NCCFS)O film is only 5°. We want to know why the water droplet cannot fall off at $\theta = 1°$ and $3°$ while easily fall off at $\theta = 5°$? This may be due to that there is a transition from Wenzel state to Cassie state when $\theta$ increases from 1° to 5° [34]. The detailed dynamic rolling process of water droplets onto the annealed (NCCFS)O film surface with $\theta = 1°$, $3°$, and $5°$ are shown in Videos 3–5 as the supplementary materials. A lower TA value of 5° indicates that the super-hydrophobic film surface owns the self-cleaning property [34].

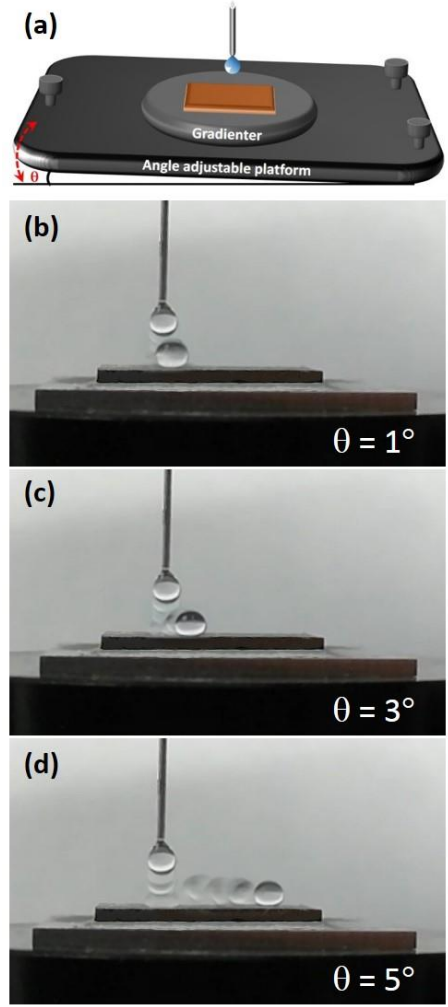

Fig. 7. Dynamic diagram of water droplets onto the annealed (NCCFS)O film surface on an inclination platform. (a) Schematic of the experimentation. $\theta$ is the tilt angle of the angle adjustable platform. Dynamic rolling process of water droplets onto the annealed (NCCFS)O film surface with (b) $\theta = 1°$, (c) $\theta = 3°$, and (d) $\theta = 5°$.

## IV. CONCLUSION

In summary, an annealed (NiCuCrFeSi)O ((NCCFS)O) high-entropy film can serve as a mid-infrared emitter due to its three characteristics: i) High blackbody-like emission in a broadband mid-infrared region; ii) High emittance in a wide temperature range; iii) Super-hydrophobicity with low title angle. These characteristics originate from the recrystallization of as-deposited film and the formation of a cavernous structure with a plethora of nanoscale polyhedrons and holes after the annealing process, supported by the results of XRD and SEM/EDX. This work may provide fresh ideas for applications of thermal emitters operating at substantially wide temperature range.

**ACKNOWLEDGMENTS**

The authors would like to acknowledge the financial support by the National Natural Science Foundation of China (NSFC) (Grants No. 51472017, No. 51732001, No. 51572010, and No. U1832219), the Fundamental Research Funds for the Central Universities, the Aeronautical Science Foundation of China, the program of China Scholarships Council (No. 201806020161), and the Academic Excellence Foundation of BUAA for PhD Students.